\documentclass[sigconf,natbib=true]{acmart}
\setcopyright{none}
\acmConference[SIGIR '26]{Proceedings of the 49th International ACM SIGIR Conference on Research and Development in Information Retrieval}{July 20--24, 2026}{Melbourne, VIC, Australia}
\acmBooktitle{Proceedings of the 49th International ACM SIGIR Conference on Research and Development in Information Retrieval (SIGIR '26), July 20--24, 2026, Melbourne, VIC, Australia}
\acmISBN{}
\acmDOI{}
\renewcommand\footnotetextcopyrightpermission[1]{}

\usepackage{multirow}

\begin{document}
\title{JFinTEB: Japanese Financial Text Embedding Benchmark}
\author{Masahiro Suzuki}
\email{research@msuzuki.me}
\orcid{0000-0001-8519-5617}
\affiliation{
  \institution{Amova Asset Management Co., Ltd.}
  \city{Tokyo}
  \country{Japan}
}
\author{Hiroki Sakaji}
\email{sakaji@ist.hokudai.ac.jp}
\orcid{0000-0001-5030-625X}
\affiliation{
  \institution{Hokkaido University}
  \city{Hokkaido}
  \country{Japan}
}

\newcommand{\subscript}[1]{$\,_\mathrm{#1}$}

\begin{abstract}
  We introduce JFinTEB, the first comprehensive benchmark specifically designed for evaluating Japanese financial text embeddings.
  Existing embedding benchmarks provide limited coverage of language-specific and domain-specific aspects found in Japanese financial texts.
  Our benchmark encompasses diverse task categories including retrieval and classification tasks that reflect realistic and well-defined financial text processing scenarios.
  The retrieval tasks leverage instruction-following datasets and financial text generation queries, while classification tasks cover sentiment analysis, document categorization, and domain-specific classification challenges derived from economic survey data.
  We conduct extensive evaluations across a wide range of embedding models, including Japanese-specific models of various sizes, multilingual models, and commercial embedding services.
  We publicly release JFinTEB datasets and evaluation framework at \url{https://github.com/retarfi/JFinTEB} to facilitate future research and provide a standardized evaluation protocol for the Japanese financial text mining community.
  This work addresses a critical gap in Japanese financial text processing resources and establishes a foundation for advancing domain-specific embedding research.
\end{abstract}


\maketitle

\section{Introduction}

\begin{figure}
  \centering
  \includegraphics[width=\linewidth]{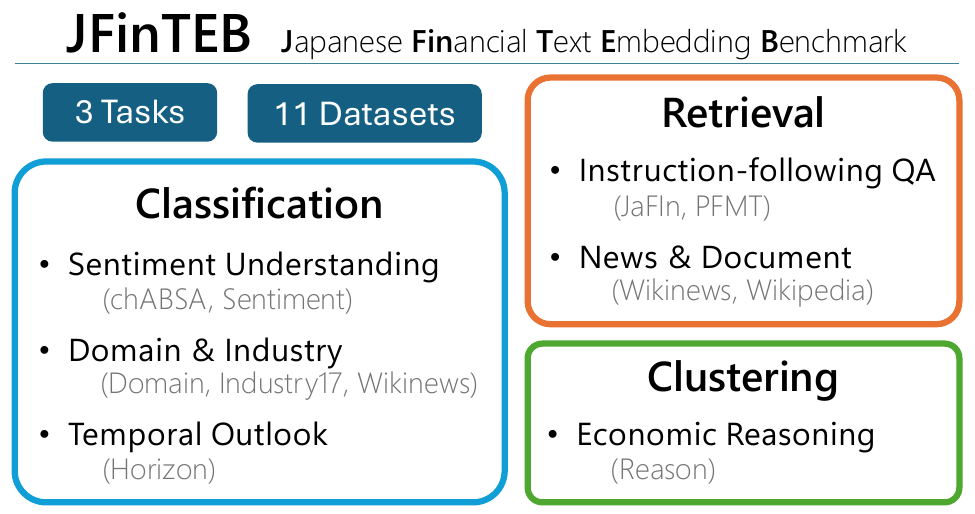}
  \caption{Overview of JFinTEB benchmark. JFinTEB is a comprehensive benchmark for evaluating Japanese financial text embeddings, covering diverse tasks such as classification, retrieval, and clustering.
  }
  \label{fig:overview}
\end{figure}

Text embeddings provide a unified representation for textual data and underpin a wide range of text mining and natural language processing tasks.
Embedding-based retrieval is a core component of modern information retrieval systems, including search, question answering, and recommendation.
The development of comprehensive benchmarks has been crucial for advancing embedding quality, with established evaluations including MTEB for English~\cite{mteb}, MMTEB for multilingual contexts~\cite{mmteb}, and JMTEB for Japanese~\cite{jmteb}.
Recently, domain-specific benchmarks such as FinMTEB have demonstrated the importance of specialized evaluation for financial applications in English and Chinese~\cite{finmteb}.

Despite Japan's position as a major financial market and the growing adoption of NLP technologies in Japanese financial institutions, no unified benchmark exists for evaluating Japanese financial text embeddings.
This gap represents a significant limitation for developing and deploying embedding models in Japan's financial sector, where domain-specific language patterns, regulatory terminology, and cultural contexts require specialized evaluation.

In particular, Japanese financial texts exhibit characteristics that make general-purpose benchmarks insufficient. 
Financial disclosures such as quarterly reports and securities filings use highly domain-specific terminology and formulaic phrasing that rarely appear in general Japanese corpora. 
Such linguistic phenomena often arise in scenarios where embeddings are directly used for retrieval, clustering, or zero-shot classification—settings common in financial information systems but underrepresented in existing benchmarks. 
These considerations underscore the necessity of a dedicated benchmark to systematically evaluate embedding models for Japanese finance.

Existing benchmarks face limitations when applied to Japanese financial contexts. 
JMTEB~\cite{jmteb} focuses on general Japanese tasks but does not include financial applications, while FinMTEB~\cite{finmteb} targets the financial domain but only in English and Chinese. 
Furthermore, financial text processing demands specialized capabilities including regulatory document analysis, market sentiment understanding, and industry classification—requirements not adequately addressed by current evaluation frameworks.
Several related efforts have proposed financial NLP tasks and evaluations.
NTCIR U4~\cite{kimura2025u4} addresses Japanese financial document understanding through table retrieval and extraction, and Hirano~\cite{hirano2024} studies the performance of large language models on Japanese financial tasks.
NTCIR FinNum~\cite{chen2018numeral} investigates numerical semantics in financial texts in English.
While these works provide valuable task-specific insights, they are not designed to evaluate text embeddings under a unified, multi-task benchmarking framework.

To address these limitations, we introduce JFinTEB (Japanese Financial Text Embedding Benchmark), the first comprehensive benchmark for evaluating Japanese financial text embeddings.
Figure~\ref{fig:overview} illustrates the overall structure of our benchmark.
Our benchmark comprises 11 carefully designed tasks spanning classification, retrieval, and clustering across diverse financial contexts, from regulatory documents to market sentiment analysis.
We establish rigorous quality assurance protocols and evaluate 14 representative embedding models, providing baseline results and practical insights for model selection in Japanese financial applications.
As a result, JFinTEB complements existing benchmarks by introducing financial-domain tasks in Japanese, filling a critical gap for both domestic applications and cross-lingual evaluation.
Table~\ref{tab:benchmark_comparison} summarizes their scope compared to JFinTEB, highlighting how our resource complements prior work. 
Unlike prior benchmarks, JFinTEB focuses on the Japanese financial domain, which combines language-specific challenges with application scenarios that are critical for industry but missing in existing resources.
Rather than constructing artificially difficult tasks, JFinTEB focuses on realistic and well-defined information needs observed in Japanese financial applications, where embeddings are commonly used for retrieval, clustering, and classification.
Consequently, some retrieval tasks exhibit high performance under current models, highlighting the maturity of embedding methods in realistic financial settings.
To our knowledge, JFinTEB is the first benchmark that systematically evaluates Japanese financial text embeddings across multiple tasks under a unified evaluation protocol.
Our design enables reproducible evaluation of text embeddings under data distributions commonly encountered in practical information retrieval systems.

The contributions of this work are threefold: (1) development of the first comprehensive Japanese financial text embedding benchmark with 11 validated tasks; (2) systematic evaluation of 14 embedding models including Japanese-specialized and multilingual approaches; and (3) public release of all datasets, evaluation code, and baseline results at \url{https://github.com/retarfi/JFinTEB}, to facilitate reproducible research in Japanese financial text mining.

{\tabcolsep=1.0pt
\begin{table}[t]
\centering
\caption{
Comparison with prior embedding benchmarks.
Abbreviations: Cls = Classification, Ret = Retrieval, Clus = Clustering, 
STS = Semantic Textual Similarity, RR = Reranking, PC = Pair Classification,
Summ = Summarization, BM = Bitext Mining.
}
\begin{tabular}{lccc}
\toprule
Benchmark & Lang. & Domain & Tasks \\
\midrule
MTEB & En & General & Cls, Ret, STS, RR, PC, Summ, BM \\
JMTEB & Ja & General & Cls, Ret, STS, RR, PC \\
FinMTEB & En, Zh & Finance & Cls, Ret, Clus, RR, PC, Summ \\
\textbf{JFinTEB (ours)} & Ja & Finance & Cls, Ret, Clus \\
\bottomrule
\end{tabular}
\label{tab:benchmark_comparison}
\end{table}
}

\section{JFinTEB Benchmark}
\subsection{Task Design and Dataset Construction}
\subsubsection{Classification Tasks}
We include several existing datasets: chABSA for aspect-based sentiment~\cite{chabsa}, three tasks derived from the Economy Watchers Survey (domain, sentiment, horizon)~\cite{suzuki2025www}, MultiFin-ja for financial news headlines~\cite{jorgensen-etal-2023-multifin}, and Wikinews classification~\cite{nishikawa-etal-2022-ease}.
``Horizon'' performs binary classification of EWS comments into current-state and future-outlook categories, utilizing the inherent survey structure that collects assessments for both present and prospective economic conditions. 

In addition, we construct two new datasets: Industry 17 and Industry 33, where company descriptions from Japanese Wikipedia are aligned with official JPX industry categories at two granularities (17 and 33 sectors), enabling coarse- and fine-grained evaluation.
Company pages of Tokyo Stock Exchange Prime Market listed firms are identified by parsing the listing information template in Japanese Wikipedia articles and extracting stock codes via automated matching with the JPX database, with no manual verification.
Since JPX industry classifications assign each company to exactly one sector, no ambiguity arises in label assignment.
Industry labels follow the official classification published by Japan Exchange Group (JPX)~\footnote{\url{https://www.jpx.co.jp/english/markets/statistics-equities/misc/01.html}}.
Wikinews classification categorizes Wikinews articles into politics and economics domains, extracted from the broader categorical structure used in previous studies~\cite{nishikawa-etal-2022-ease}.

\subsubsection{Retrieval Tasks}
Our four retrieval tasks cover diverse information access needs. 
JaFIn~\cite{jafin} evaluates retrieval of financial FAQs, and PFMT~\cite{Hirano2025-nlp} provides a multi-turn regulatory Q\&A benchmark. 
In addition, we construct two retrieval datasets using automated procedures with no manual annotation.
Wikinews retrieval matches news headlines with their corresponding articles, drawn from the same politics and economics categories described above.
Wikipedia retrieval pairs company names --- taken directly from Japanese Wikipedia article titles --- with their corresponding descriptions, using the same set of Tokyo Stock Exchange Prime Market listed companies identified for Industry 17/33.
These additions ensure coverage of news-driven and corporate information scenarios specific to the Japanese financial context.

\subsubsection{Clustering Tasks}
We extend prior setups~\cite{suzuki2025www} by introducing ``Reason'', which groups comments into 13 economic reasoning categories. 
Unlike earlier work that included an ``other'' class, we redefine two frequent categories (trends in hires, employment type characteristics) to provide more balanced clustering.

\subsection{Dataset Statistics and Availability}
Table~\ref{tab:task-stats} presents statistics for all tasks.
Retrieval and clustering tasks use validation sets only for selecting evaluation configurations, with no model training; hence, the absence of training sets.
All datasets curated in this study (Horizon, Wikinews (classification and retrieval), Industry 17, Industry 33, and Wikipedia-retrieval) are publicly available at \url{https://github.com/retarfi/JFinTEB} and contain no personally identifiable information.

\begin{table}[tb]
  \begin{center}
  \caption{
    JFinTEB Task Statistics.
    Chars indicates the median number of characters in the validation (val.) set.
  }
  \label{tab:task-stats}
  \begin{tabular}{lccccc} \toprule
    Name & Train & Val. & Test & Chars & JFinTEB \\
    \midrule
    \textbf{Classification} \\
    chABSA & 1558 & 222 & 447 & 72 & $\checkmark$ \\
    Domain & 4200 & 600 & 1200 & 52 & $\checkmark$ \\
    Sentiment & 7000 & 1000 & 2000 & 51 & $\checkmark$ \\
    Horizon & 7000 & 1000 & 2000 & 51 & $\checkmark$ \\
    Wikinews & 683 & 97 & 196 & 526 & $\checkmark$ \\
    Industry 17 & 1034 & 147 & 297 & 298 & $\checkmark$ \\
    Industry 33 & 1034 & 147 & 297 & 298 & \\
    MultiFin-ja & 147 & 37 & 46 & 53 & \\
    \midrule
    \textbf{Retrieval} \\
    JaFIn & - & 298 & 1192 & 184 & $\checkmark$ \\
    PFMT & - & 60 & 240 & 825 & $\checkmark$ \\
    Wikinews & - & 199 & 798 & 551 & $\checkmark$ \\
    Wikipedia & - & 295 & 1183 & 327 & $\checkmark$ \\
    \midrule
    \textbf{Clustering} \\
    Reason & - & 6500 & 6500 & 53 & $\checkmark$ \\
    \bottomrule
  \end{tabular}
  \end{center}
\end{table}

\subsection{Evaluation Methodology}
Following JMTEB~\cite{jmteb}, we adopt standard evaluation protocols for classification (macro-F1), retrieval (NDCG@10), and clustering (V-measure). 
Our implementation builds directly on the JMTEB codebase with minor modifications to incorporate financial datasets, ensuring consistency and reproducibility across benchmarks.
For all tasks, validation sets are used exclusively to select evaluation configurations, while test sets are held out for final reporting.
For retrieval and clustering tasks, no model training is performed; validation data are used only to select evaluation settings, ensuring fair and reproducible comparisons across models.

\subsection{Quality Assurance and Validation}
We validate task quality using two stability criteria to ensure reliable evaluation:

\noindent \textbf{Model Family Consistency:} Using three embedding model families with different parameter scales---Multilingual E5 (small/large)~\cite{multilinguale5}, Ruri v3 (30M/310M)~\cite{tsukagoshi2024ruri}, and OpenAI text-embedding-3 (small/large)---we identify tasks showing size-performance reversals across two or more families.
Only MultiFin-ja exhibited such reversals across two families (E5 and OpenAI), likely due to its significantly smaller sample size as shown in Table~\ref{tab:task-stats}.

\noindent \textbf{Validation-Test Stability:} We exclude tasks with validation-test score differences exceeding 20\% in the three large models, indicating potential distribution mismatches or evaluation instabilities.
One classification task (Industry 33) showed excessive validation-test divergence in the three models.

Based on these criteria, we exclude MultiFin-ja and Industry 33 from the final benchmark.
The resulting JFinTEB comprises 11 stable tasks across classification, retrieval, and clustering, ensuring consistent and interpretable evaluation results across diverse embedding models.

\section{Evaluation}
We evaluate representative embedding models across different architectures and languages, primarily selected based on strong performance reported in JMTEB.
Table~\ref{tab:models} summarizes the model statistics, including parameter sizes and maximum input lengths.

\noindent \textbf{Japanese-Specialized Models} We evaluate leading Japanese embedding models: (1) Ruri v3 series (Ruri)~\cite{tsukagoshi2024ruri}, trained with contrastive learning and based on Japanese ModernBERT~\cite{modernbert-ja}; (2) Sarashina~\footnote{\url{https://huggingface.co/sbintuitions/sarashina-embedding-v1-1b}}, which is derived from a 1.2B Japanese LLM with multi-stage training and achieves state-of-the-art JMTEB performance; and (3) GLuCoSE~\footnote{\url{https://huggingface.co/pkshatech/GLuCoSE-base-ja-v2}}, a LUKE-based~\cite{yamada2020luke} model optimized for Japanese semantic tasks.

\noindent \textbf{Multilingual Models}
We include three high-performing multilingual embedding families: (1) \texttt{jina-embeddings-v3}~\cite{jina-embeddings-v3} (Jina), a multi-task model based on XLM-RoBERTa~\cite{conneau-etal-2020-unsupervised} with 8192-token capacity and LoRA~\cite{hu2022lora} adapters; (2) \texttt{Multilingual E5} series~\cite{multilinguale5} (E5); and (3) \texttt{OpenAI text-embedding-3} (OpenAI).

\noindent \textbf{Domain Adaptation Baselines} We include Japanese BERT (BERT)~\footnote{\url{https://huggingface.co/tohoku-nlp/bert-base-japanese}} and Japanese financial BERT (FinBERT) \cite{Suzuki-2023-ipm}, which is additionally pre-trained on financial corpora, to assess financial domain adaptation effectiveness.

{\tabcolsep=3.5pt
\begin{table}[tb]
  \begin{center}
  \caption{Evaluated Model Statistics}
  \label{tab:models}
  \begin{tabular}{lcc}
    \toprule
    Model & Size & Max Tokens\\
    \midrule
    \multicolumn{3}{l}{\textbf{Japanese-Specialized Models}} \\
    Sarashina & 1.22B & 8192 \\
    Ruri\subscript{310M/130M/70M/30M} & 310M, 130M, 70M, 30M & 8192 \\
    GLuCoSE & 133M & 512 \\
    \midrule
    \multicolumn{3}{l}{\textbf{Multilingual Models}} \\
    Jina & 572M & 8192 \\
    E5\subscript{large/base/small} & 560M, 278M, 118M & 512 \\
    OpenAI\subscript{large/small} & - & 8192 \\
    \midrule
    \multicolumn{3}{l}{\textbf{BERT Models}} \\
    BERT & 111M & 512 \\
    FinBERT & 111M & 512 \\
    \bottomrule
  \end{tabular}
  \end{center}
\end{table}
}

{\tabcolsep=3.5pt
\begin{table*}[tb]
  \begin{center}
  \caption{Evaluation Results on JFinTEB Tasks.
  Avg. is the average score across all the tasks (excluding MultiFin-ja and Industry 33.)
  Best scores are in \textbf{bold}.
  Clust. refers to a clustering task.
  }
  \label{tab:result}
  \begin{tabular}{l*{13}{c}}
    \toprule
     & \multicolumn{6}{c}{Classification} & \multicolumn{4}{c}{Retrieval} & Clust. & \multirow{2}{*}{Avg.} \\ \cmidrule(lr){2-7} \cmidrule(lr){8-11} \cmidrule(lr){12-12}
    Model & chABSA & Domain & Horizon & Sentiment & Wikinews & Industry 17 & JaFIn & PFMT & Wikinews & Wikipedia & Reason & \\
    \midrule
    \multicolumn{12}{l}{\textbf{Japanese-Specialized Models}} \\
    Sarashina & \textbf{96.0} & 78.1 & 82.8 & 55.9 & 94.6 & \textbf{82.9} & 86.3 & 95.5 & 93.3 & \textbf{98.7} & 26.2 & \textbf{80.9} \\
    Ruri\textsubscript{310M} & 95.6 & 77.1 & 83.0 & 53.9 & 94.3 & 72.2 & 85.3 & 95.7 & 93.1 & 96.7 & 26.4 & 79.4 \\
    Ruri\textsubscript{130M} & 90.3 & 77.7 & 82.6 & 54.9 & 94.6 & 75.8 & 84.1 & 95.4 & \textbf{93.5} & 95.9 & 24.8 & 79.1 \\
    Ruri\textsubscript{70M} & 92.6 & 76.9 & 81.5 & 52.1 & 93.9 & 72.9 & 82.1 & 95.2 & 93.1 & 91.9 & 22.3 & 77.7 \\
    Ruri\textsubscript{30M} & 91.1 & 76.9 & 80.0 & 51.5 & 94.1 & 77.5 & 81.2 & 93.7 & 92.9 & 94.1 & 24.8 & 78.0 \\
    GLuCoSE & 92.8 & 76.4 & 80.7 & 49.8 & 95.0 & 79.1 & 78.3 & 97.8 & 91.7 & 96.8 & 25.7 & 78.6 \\
    \midrule
    \multicolumn{12}{l}{\textbf{Multilingual Models}} \\
    Jina & 92.5 & 77.3 & 83.0 & 51.0 & 95.6 & 65.1 & \textbf{86.4} & \textbf{98.5} & 92.7 & 98.5 & 28.6 & 79.0 \\
    E5\textsubscript{large} & 93.9 & 74.6 & 81.9 & 53.3 & 95.2 & 79.1 & 81.8 & 95.6 & 91.5 & 98.0 & 27.2 & 79.3 \\
    E5\textsubscript{base} & 92.1 & 77.4 & 81.7 & 50.3 & 94.8 & 71.9 & 70.5 & 94.3 & 88.7 & 93.3 & 23.1 & 76.2 \\
    E5\textsubscript{small} & 90.4 & 77.9 & 77.5 & 47.0 & \textbf{96.0} & 69.9 & 66.9 & 96.8 & 84.3 & 93.7 & 26.6 & 75.2 \\
    OpenAI\textsubscript{large} & 93.4 & \textbf{80.6} & \textbf{84.5} & \textbf{57.5} & 94.1 & 58.4 & 84.3 & 95.9 & 92.9 & 97.5 & \textbf{29.6} & 79.0 \\
    OpenAI\textsubscript{small} & 87.5 & 78.8 & 80.9 & 52.1 & 91.3 & 51.5 & 76.0 & 94.9 & 89.7 & 87.5 & 27.4 & 74.3 \\
    \midrule
    \multicolumn{12}{l}{\textbf{BERT Models}} \\
    BERT & 93.0 & 73.6 & 82.0 & 51.3 & 93.5 & 76.5 & 31.6 & 40.8 & 57.4 & 14.1 & 15.6 & 57.2 \\
    FinBERT & 94.9 & 75.4 & 81.5 & 51.0 & 93.9 & 77.7 & 35.3 & 50.0 & 52.7 & 21.8 & 18.5 & 59.3 \\
    \bottomrule
  \end{tabular}
  \end{center}
\end{table*}
}

\section{Discussion}
Table~\ref{tab:result} shows the evaluation results on JFinTEB tasks.

\noindent \textbf{Japanese vs Multilingual Models}
Japanese-specialized models achieved average scores of 80.9 (Sarashina) and 79.4 (Ruri\textsubscript{310M}), while multilingual models scored 79.3 (E5\textsubscript{large}) and 79.0 (Jina, \linebreak OpenAI\textsubscript{large}).
This performance gap is comparable to results reported in JMTEB for general Japanese language tasks.
While Japanese-specialized models maintain advantages across domains, the results also suggest significant improvements in Japanese language capabilities of multilingual models.

\noindent \textbf{Task-Specific Performance}
Computing standard deviations of performance differences between Japanese-specialized and multilingual models reveals the highest variance in JaFIn (0.063), followed by Wikipedia retrieval (0.033), Sentiment (0.029), and Wikinews retrieval (0.027), indicating that retrieval tasks show larger performance variations.
Given that classification tasks involve fine-tuning, retrieval tasks may better reflect models' generalization capabilities.
Averaging performance within task categories, classification tasks rank models as: Sarashina (81.7), E5\textsubscript{large} (79.7), Ruri\textsubscript{310M} (79.4), and GLuCoSE (79.0), while retrieval tasks rank them as: Jina (94.0), Sarashina (93.5), and Ruri\textsubscript{310M}/OpenAI\textsubscript{large} (92.7).
Although the best-performing model varies by task, Sarashina, which achieved the highest overall average, ranks first in classification categories and second in retrieval categories, demonstrating consistently strong performance across diverse tasks.
OpenAI\textsubscript{large} achieves the highest performance on Economy Watchers Survey-derived tasks (Domain, Horizon, Sentiment, Reasoning), with OpenAI\textsubscript{small} also showing relatively strong performance on these tasks compared to others.
This suggests that OpenAI models were likely trained on Economy Watchers Survey texts.
Beyond task-specific training, broad corpus training appears crucial for achieving generalizable performance.

\noindent \textbf{Retrieval Task Saturation}
High retrieval scores suggest that several retrieval tasks reflect well-defined information needs commonly observed in real-world financial information retrieval scenarios.
Such saturation does not indicate task triviality, but rather highlights the maturity of current embedding models under realistic data distributions.
At the same time, these results motivate future extensions toward harder retrieval settings, such as longer documents or cross-document reasoning, to further stress embedding capabilities.

\noindent \textbf{Model Size}
Both Japanese-specialized and multilingual model families demonstrate consistent trends where larger models achieve higher performance.
Considering overall performance, Sarashina represents the optimal choice for models exceeding 1B parameters, while Ruri\textsubscript{310M} is a strong candidate for 100M-1B parameter range.
For computational constraints below 100M parameters, OpenAI\textsubscript{large} via API is preferable when accessible, otherwise Ruri\textsubscript{30M} provides the best alternative.

\noindent \textbf{Domain Adaptation}
Comparing BERT-family models, FinBERT trained on financial corpora outperforms general-purpose BERT.
The effectiveness of adapting general embedding models to financial domains has been demonstrated for English and Chinese~\cite{finmteb}, and similar domain-adapted embedding models for Japanese—a relatively minor language compared to English—are expected to be developed, demonstrating domain adaptation benefits.

\noindent \textbf{Data Contamination Considerations}
Most datasets in JFinTEB are derived from long-standing publicly available sources, such as Wikipedia, Wikinews, and the Economy Watchers Survey.
As a result, strong performance by large commercial models may partly reflect exposure during pretraining rather than purely task-level generalization.
Rather than eliminating such datasets, JFinTEB supports analysis of embedding robustness under realistic data availability conditions, which is critical for information retrieval systems deployed in practice.

\section{Resource Availability and Ethics}
All datasets, evaluation scripts, and baseline results of JFinTEB are publicly available.
The full benchmark implementation, including dataset loaders, evaluation pipelines,
and baseline scripts, is hosted on GitHub\footnote{\url{https://github.com/retarfi/JFinTEB}}.
JFinTEB includes newly curated datasets based on Wikipedia and Wikinews, while other benchmark tasks rely on publicly available external datasets that are automatically retrieved from the HuggingFace Hub\footnote{\url{https://huggingface.co/datasets/retarfi/JFinTEB}}.
Each dataset is used under its original license or terms of use.

The benchmark does not contain personally identifiable information.
All textual data are obtained from publicly released documents or anonymized survey responses, and no human subjects experiments were conducted for constructing JFinTEB.

We acknowledge potential biases inherent in the data sources.
For example, the Economy Watchers Survey reflects perspectives of specific respondent groups, and Wikipedia-based datasets may exhibit coverage biases.

We plan to maintain JFinTEB by addressing issues reported by users and by incorporating new tasks as Japanese financial NLP applications evolve.

\section{Conclusions}
We present JFinTEB, the first comprehensive benchmark for evaluating Japanese financial text embeddings, comprising 11 validated tasks across classification, retrieval, and clustering. 
Our evaluation of 14 embedding models reveals several key insights.
Japanese-specialized models achieve slight advantages over multilingual counterparts, though the gap is narrowing as multilingual models improve their Japanese capabilities.
Model size consistently correlates with performance across both Japanese-specialized and multilingual families, with Sarashina emerging as the top choice for resource-rich scenarios and Ruri v3 providing strong alternatives for computational constraints.

The benchmark identifies clear opportunities for future development.
Domain adaptation shows promise, as evidenced by FinBERT's improvements over general BERT, suggesting potential for developing Japanese financial embedding models.
We release datasets, evaluation code, and baseline results to facilitate reproducible research in Japanese financial text mining.
JFinTEB provides a foundation for advancing Japanese financial text understanding and enabling applications in Japan's financial industry.
At the same time, our current benchmark is limited to relatively short-text tasks and automatic evaluation metrics. 
Future extensions will incorporate long-document scenarios (e.g., financial reports), generative or RAG-based tasks, and human-centered evaluations to better capture the full range of financial NLP applications.
Beyond the Japanese financial domain, JFinTEB also serves as a resource for studying domain adaptation and cross-lingual benchmarking.

\begin{acks}
This work was supported by JST PRESTO, Japan, Grant Number JPMJPR2267.
The views expressed in this paper are solely those of the author and do not represent the official position of Amova Asset Management.
\end{acks}

\bibliographystyle{ACM-Reference-Format}
\bibliography{reference}

\end{document}